\documentclass[a4paper,11pt]{article}
\usepackage{jheppub}
\usepackage{orcidlink}

\usepackage{subfigure}
\usepackage{mathtools}
\usepackage{amsmath}
\usepackage{amsfonts}
\usepackage{amssymb}
\usepackage{bbold}
\usepackage{bm}
\usepackage{ esint }
\usepackage{orcidlink}

\newcommand{\br}{{\bf r}}

\newcommand{\bz}{{\bf z}}

\newcommand{\bx}{{\bf x}}
\newcommand{\bn}{{\bf n}}
\newcommand{\bk}{{\bf k}}
\newcommand{\bK}{{\bf K}}

\newcommand{\bv}{{\bf v}}
\newcommand{\bp}{{\bf p}}
\newcommand{\bpartial}{\boldsymbol{\partial}}

\newcommand{\bu}{{\bf u}}

\newcommand{\bG}{{\bf G}}

\newcommand{\sOm}{{\sf\Omega}}

\newcommand{\bpsi}{{\boldsymbol{\psi}}}

\long\def\exclude#1{}

\newcommand{\GF}{G_{\rm F}}

\newcommand{\uR}{u_{\rm R}}
\newcommand{\uI}{u_{\rm I}}

\title{Theory of neutrino fast flavor evolution.\\
Part II. Solutions at the edge of instability.}

\author[a]{Damiano F.\ G.\ Fiorillo \orcidlink{0000-0003-4927-9850}} 
\affiliation[a]{Deutsches Elektronen-Synchrotron DESY,
Platanenallee 6, 15738 Zeuthen, Germany}

\author[b]{and Georg G.\ Raffelt
\orcidlink{0000-0002-0199-9560}}
\affiliation[b]{Max-Planck-Institut f\"ur Physik, Boltzmannstr.~8, 85748 Garching, Germany}

\abstract{In dense neutrino environments, such as provided by core-collapse supernovae or neutron-star mergers, neutrino angular distributions may be unstable to collective flavor conversions, whose outcome remains to be fully understood. These conversions are much faster than hydrodynamical scales, suggesting that self-consistent configurations may never be strongly unstable. With this motivation in mind, we study weakly unstable modes, i.e., those with small growth rates. We show that our newly developed dispersion relation (Paper~I of this series) allows for an expansion in powers of the small growth rate. For weakly unstable distributions, we show that the unstable modes must either move with subluminal phase velocity, or very close to the speed of light. The instability is fed from neutrinos moving resonantly with the waves, allowing us to derive explicit expressions for the growth rate. For axisymmetric distributions, often assumed in the literature, numerical examples show the accuracy of these expressions. We also note that for the often-studied one-dimensional systems one should not forget the axial-symmetry-breaking modes, and we provide explicit expressions for the range of wavenumbers that exhibit instabilities.
}

\begin{document}
\maketitle
\flushbottom

\section{Introduction}

Flavor evolution in neutrino-dense environments, notably in core-collapse supernovae and neutron-star mergers, is characteristically of a collective nature. The weak interaction among neutrinos allows for refractive flavor exchange~\cite{Pantaleone:1992eq, Samuel:1993uw} that can induce conversions over timescales much faster than vacuum oscillations~\cite{Samuel:1995ri}. This effect is caused by run-away modes of the collective neutrino flavor field. Samuel's original bimodal instability~\cite{Samuel:1995ri} is caused by an interplay between vacuum oscillations and neutrino-neutrino refraction, but later Sawyer discovered another type of unstable collective mode that does not require neutrino masses \cite{Sawyer:2004ai, Sawyer:2008zs, Sawyer:2015dsa} and is today referred to as fast flavor instability \cite{Chakraborty:2016lct, Izaguirre:2016gsx, Tamborra:2020cul, Richers:2022zug, Patwardhan:2022mxg}. Under appropriate conditions, these conversions are so rapid that neutrino masses, originally introduced to explain vacuum flavor conversions, ironically become irrelevant; they only provide a necessary small deviation from pure flavor eigenstates, which is self-amplified purely by weak interactions. Fast flavor conversions represent the purest form of collective flavor evolution, entirely driven by neutrino-neutrino refraction.

The community has only slowly appreciated the relevance of fast flavor modes from an evolving perspective on collective flavor evolution. Sawyer's early findings~\cite{Sawyer:2004ai, Sawyer:2008zs, Sawyer:2015dsa} were based on perfectly homogeneous systems of a few discrete beams. The notion of fast conversions truly thrived when it was realized that the instability is of a kinetic nature~\cite{Izaguirre:2016gsx}, due to the interplay between self-interaction and the streaming of neutrinos. Therefore, even small inhomogeneities grow unstable, with a characteristic length scale determined by the reciprocal of the interaction strength $\mu=\sqrt{2}\GF (n_\nu+n_{\overline{\nu}})$, where $\GF$ is the Fermi constant and $n_{\nu}$ and $n_{\overline{\nu}}$ are the neutrino and antineutrino densities. While this is the commonly used definition for the interaction strength, a sometimes underemphasized point is that fast dynamics is actually sensitive only to the lepton number, so that the actual length and time scales are more correctly of the order of $\mu \epsilon$, where $\epsilon=(n_{\nu}-n_{\overline{\nu}})/(n_\nu+n_{\overline{\nu}})$. In the decade that followed, some fundamental features of fast flavor evolution were identified, mostly through either formal arguments~\cite{Morinaga:2021vmc}, or inference from numerical simulations (see, e.g., Refs.~\cite{Bhattacharyya:2020jpj, Zaizen:2022cik, Nagakura:2023jfi, Xiong:2023vcm, Shalgar:2022lvv, Cornelius:2023eop}). It is now well understood that fast instabilities require a so-called angular crossing, namely a change in sign of the angular distribution of the electron-minus-muon lepton number; this was proven formally by Morinaga~\cite{Morinaga:2021vmc} using algebraic properties of the dispersion relation. In addition, numerical simulations have shown~\cite{Nagakura:2022kic} that at least locally -- on length and time scales of order $\mu^{-1}$ -- the instability saturates by removing the angular crossing, with one of the sides of the crossing reaching equipartition. However, the extent and generality of this finding is hard to assess through numerical simulations.

As our understanding of specific features of fast conversions has grown, a consistent framework to interpret and intuitively understand them is still lacking. In the present series, starting with Paper~I~\cite{Fiorillo:2024bzm}, we push this frontier. In Paper~I, we have contextualized flavor instabilities as a new class of kinetic instabilities, analogous to the plasma ones thoroughly studied since the 1960s. The unstable growth can be understood based on the picture of flavor waves, namely spacetime periodic oscillations of the flavor coherence. Individual neutrinos can resonantly exchange energy with flavor waves moving with the same phase velocity, through a coupling proportional to the electron-minus-muon lepton number. Therefore, if the latter changes sign at an angular crossing, there will be waves resonant with neutrinos on one side of the crossing that will be amplified. Based on the resonance picture, we provided an intuitive proof of Morinaga's result~\cite{Morinaga:2021vmc}. In this framework, the growth rate of weakly unstable configurations is expected to be proportional to the amount of lepton number on one side of the crossing, although the precise relation was not clarified in Paper~I. 

This question is especially important because strongly unstable configurations are never truly expected in Nature, since they would relax very rapidly; this inconsistency was first highlighted by Johns~\cite{Johns:2023jjt,Johns:2024dbe} within his hypothesis that some form of thermalization should attain. While the latter remains speculative, the conceptual inconsistency of strongly unstable configurations is generic, and we have later shown for a simple example that in reality the system would rather move along the edge of stability~\cite{Fiorillo:2024qbl}, quickly eliminating the source of instability. In this way, we proved, using a quasi-linear approach, for a discrete set of beams the removal of the angular crossing found in numerical works. The appearance of equipartition on one side of the crossing is now easy to understand, and it has the same nature as, e.g., the removal of bumps in the velocity distribution of particles in a plasma due to the bump-on-tail instability~\cite{Vedenov1962quasi, drummond1961non}. 

Meanwhile, the renewed emphasis on weakly unstable configurations raises back a new question, namely the impact of neutrino masses, which a few recent papers have considered~\cite{Shalgar:2020xns, DedinNeto:2023ykt}. Within the framework that we construct here, we leave this question for future study, as well as any impact from the collisional term which might induce novel forms of instabilities~\cite{Johns:2021qby,Xiong:2022zqz, Liu:2023pjw, Lin:2022dek, Johns:2022yqy, Padilla-Gay:2022wck, Fiorillo:2023ajs}.

In this second paper, we complete the program set forth in Paper~I by bridging the intuitive framework with the more practical question of the growth rate of instability. We use the analytic properties of the dispersion relation, that we clarified in Paper~I, to obtain the main properties of weak instabilities; specifically, we identify the range of wavevectors that contains unstable modes, and we provide approximate expressions for their growth rates, proportional to the amount of resonant neutrinos. We also provide a fully worked-out example where we test these results. The approximate growth rates derived here play the same role as the growth rates of plasma instabilities which form the cornerstone of the theory of plasma turbulence. They provide qualitative understanding, and may serve as the basis for a nonlinear theory of the instability evolution.

We structure our work by first reviewing the dispersion relation for fast flavor evolution in Sec.~\ref{sec:dispersion_recap}, including the new insights on its analytic properties derived in Paper~I. In Sec.~\ref{sec:lepton_number_transversality}, we develop a previously unnoticed analogy between fast flavor and electromagnetic waves, based on a transversality condition originating from lepton number conservation; this allows us to provide a simplified form of the dispersion relation. In Sec.~\ref{sec:weak_instabilities_resonance}, we specifically consider weak instabilities, and provide a general discussion of their approximate growth rates and regions of instability. These results are later specialized to the case of axisymmetric angular distributions in Sec.~\ref{sec:axisymmetric_distributions}, and applied to a worked-out example in Sec.~\ref{sec:benchmark_example}. Finally, we summarize our conclusions in Sec.~\ref{sec:discussion}.

\section{Dispersion relation for fast flavor evolution}\label{sec:dispersion_recap}

In Paper~I we have formulated the linear theory of neutrino fast flavor evolution using a linear-response approach \cite{Fiorillo:2024bzm}. The main result needed from this earlier work is the fast flavor dispersion relation that was formulated somewhat differently from the previous literature, allowing for an expansion in powers of the growth rate. In this section, we briefly review these results and emphasize why they are relevant here.

The dispersion relation descends from the usual quantum kinetic equations (QKEs) \cite{Dolgov:1980cq, Rudsky, Sigl:1993ctk, Sirera:1998ia, Yamada:2000za, Vlasenko:2013fja, Volpe:2013uxl, Serreau:2014cfa, Kartavtsev:2015eva, Fiorillo:2024fnl, Fiorillo:2024wej} that describe collective flavor oscillations in terms of flavor density matrices for each momentum mode $\bp$. In the linear regime, the full three-flavor dynamics breaks into three independent two-flavor problems \cite{Airen:2018nvp}, and in each of them, the density matrices can be written in the traditional form
\begin{equation}
    \varrho_\bp
    =\frac{1}{2}\bigl({\rm Tr}\varrho_\bp+\vec{P}_{\bp}\cdot\vec{\sigma}\bigr)
    =\frac{1}{2}\begin{pmatrix}
        n_\bp+P^z_\bp & P^x_\bp-iP^y_\bp \\
        P^x_\bp+iP^y_\bp & n_\bp-P^z_\bp
    \end{pmatrix}.
\end{equation}
They are parameterized in terms of the phase-space density $n_\bp$ and conventional polarization vector $\vec{P}_\bp$. Following our previous notation~\cite{Fiorillo:2024fnl, Fiorillo:2024qbl, Fiorillo:2024bzm}, we denote by $\psi_\bp=P^x_\bp+iP^y_\bp$ the degree of flavor coherence, which vanishes if neutrinos are in pure flavor eigenstates. The density matrix is assumed to depend on space-time coordinates $x=(t,\br)$. The field of flavor coherence $\psi_\bp(x)$ is our central object of study and, in the linear regime, is the only dynamical variable, whereas $P^z_\bp(x)$ is fixed at its initial value that traditionally is called the spectrum $G_\bp$. The task at hand is to understand the solution for $\psi_\bp(x)$ for a given $G_\bp$ and a given spectrum of initial perturbations for $\psi_\bp(x)$.

Within our approximations, the QKEs can be regarded the collisionless kinetic equations for massless neutrinos with momentum $\bp$ and renormalized dispersion relation \cite{Fiorillo:2024fnl}
\begin{equation}
    {\sOm_\bp}(x)=|\bp|+\sqrt{2}\GF \int \frac{d^3\bp'}{(2\pi)^3}\,
    v\cdot v'\,\varrho_{\bp'}(x).
\end{equation}
Here $\GF$ is Fermi's constant, $v^\mu=p^\mu/|\bp|$ the velocity four-vector, $p^\mu=(|\bp|,\bp)$ the four-momentum, and we denote by a dot the covariant four-product. Focusing on the trace-free part of the QKE, we set $\sqrt{2}\GF=1$, which is obtained by an appropriate fixing of the units of space and time, and we consider polarization vectors integrated over energy, $\vec{P}_\bv=\int \frac{|\bp|^2 d|\bp|}{(2\pi)^3}\vec{P}_\bp$, where $\bv=\bp/|\bp|$ is the neutrino velocity, a unit vector of direction. The integration over energy is possible because energy enters the QKE only through the vacuum oscillation frequency that we neglect in the fast flavor limit. We then find
\begin{equation}\label{eq:precession-equation}
   (\partial_t+\bv\cdot\bpartial_\br)\vec{P}_\bv=
    v\cdot\partial \vec{P}_\bv=\int d^2\bv'\, v\cdot v'\,
    \vec{P}_{\bv'}\times \vec{P}_\bv;
\end{equation}
these equations conserve the lengths of the polarization vectors $\vec{P}_\bv$.

We have not yet mentioned antineutrinos. As abundantly discussed in the previous literature, the QKEs for neutrinos and antineutrinos can be combined to an equation for particle number (neutrinos plus antineutrinos) and one for lepton number (neutrinos minus antineutrinos). In the fast flavor limit, the equation for lepton number decouples from that for particle number, forming a closed system by itself. The QKE for lepton number coincides with Eq.~\eqref{eq:precession-equation}, assuming all quantities refer to flavor lepton number stored in neutrinos. The exact impact of lifting this degeneracy by the inclusion of nonvanishing vacuum oscillation frequencies remains to be investigated, or rather, what is the exact criteria for adopting the fast flavor limit.

In the instability picture, neutrinos and antineutrinos are considered to be initially in their flavor eigenstates apart from small initial seeds that one imagines descend from the otherwise neglected mass term, so the off-diagonal component $\psi_\bv$ is very small and can be treated perturbatively. The QKEs for $\psi_\bv$ then read in compact form
\begin{equation}\label{eq:EOM-linear}
    v\cdot\partial\psi_\bv=i(G\cdot v\, \psi_\bv-\psi\cdot v\, G_\bv),
\end{equation}
where we have introduced the notation 
\begin{equation}
    G^\mu=\int d^2\bv\,G_\bv v^\mu
    \quad\hbox{and}\quad
    \psi^\mu=\int d^2\bv\,\psi_\bv v^\mu,
\end{equation}
where the time-like component is the original spectrum and field of flavor coherence, whereas the space-like components are the corresponding fluxes.

Since the length scales over which flavor conversions activate are usually much shorter than the hydrodynamical length scales over which density and temperature change in SNe, we consider the neutrino angular distribution $G_\bv$ to be homogeneous. Therefore, in the conventional search for normal modes for $\psi_\bv$, one looks for solutions of the form $\psi_\bv(t,\br)=\psi_\bv(\Omega,\bK)\,e^{-i\Omega t+i\bK\cdot \br}$, where $K^\mu=(\Omega,\bK)$ is the four-dimensional physical wavevector. Moreover, we introduce the usual shifted wavevector $k^\mu=K^\mu+G^\mu$, leading immediately to the dispersion relation for normal modes
\begin{equation}\label{eq:wrong_chi}
    \psi^\mu={\chi}^{\mu\nu} \psi_\nu
    \quad\hbox{with}\quad
    {\chi}^{\mu\nu}=\int d^2\bv \frac{G_\bv v^\mu v^\nu }{k\cdot v}.
\end{equation}
It is worth noting that the often-studied homogeneous mode, defined by vanishing $\bk$, is not physically homogeneous in that it has the physical wavevector $\bK=-\bG$ unless one works in a coordinate frame where the lepton-number flux exactly vanishes.

In our companion paper~\cite{Fiorillo:2024bzm}, we have shown that a slightly different dispersion relation can be derived from linear-response theory, i.e., by considering the collective field $\psi^\mu$ as an external field. This modified dispersion relation emerges if we assume that the field is slowly applied from an infinitely remote past, so that the tensor $\chi^{\mu\nu}$ acquires the physical role of a flavor susceptibility and is modified to
\begin{equation}
    {\chi}^{\mu\nu}=\int d^2\bv \frac{G_\bv v^\mu v^\nu }{k\cdot v+i\epsilon},
\end{equation}
where $\epsilon$ is an infinitely small positive number. This is a definite prescription to avoid the pole $k\cdot v=0$, the latter having the physical meaning of resonant emission or absorption of Cherenkov flavor waves from individual neutrinos \cite{Fiorillo:2024bzm}. The modified dispersion relation
\begin{equation}\label{eq:dispersion_relation_4d}
    \mathrm{det}(g^{\mu\nu}-\chi^{\mu\nu})=0
\end{equation}
does not provide the normal modes of the system, but rather its late-time asymptotic behavior.

For growing modes, such that $\mathrm{Im}(\Omega)>0$, the two alternative definitions of $\chi^{\mu\nu}$ lead exactly to the same dispersion relation, so that the more conventional search for growing modes is not affected. However, when $\mathrm{Im}(\Omega)\to 0$, the modified dispersion relation has a definite practical advantage, because it is an analytic function of the complex frequency $\Omega$. This means that it can be differentiated for small growth rates. As we will see, this is the key property that makes it more instructive to determine the properties of systems close to stability, for which $\mathrm{Im}(\Omega)$ is small.

Finally, in our companion paper~\cite{Fiorillo:2024bzm} we have also shown that the solution can be alternatively formulated in terms of the complex phase velocity $u=\omega/|\bk|$, rather than the frequency $\omega$ itself. If we define $\bn=\bk/|\bk|$ as the unit vector in the direction of the wavevector $\bk$, and $\kappa=|\bk|$, we can rewrite the dispersion relation as
\begin{equation}\label{eq:dispersion_kappa}
    \Phi(u,\kappa,\bn)=\mathrm{det}(\kappa \delta^{\mu\nu}-\tilde{\chi}^{\mu\nu})=0,
\end{equation}
where
\begin{equation}
    \tilde{\chi}^{\mu\nu}=\int d^2\bv\,\frac{G_\bv v^\mu v^\nu}{u-\bn\cdot\bv+i\epsilon}.
\end{equation}
It is then clear that for each value of $\bu$ the module of the wavevector can be found as the eigenvalue of the matrix $\tilde{\chi}^\mu_\nu$. In this formulation, the dispersion relation is not just a self-consistency condition, but rather an eigenvalue equation to find $\kappa$. Notice that this is an eigenvalue for a pseudo-Euclidean matrix, and therefore, even if $\chi^{\mu\nu}$ were real, its symmetric nature does not guarantee that all of its eigenvalues are real; see our discussion in  Sec.~6.2 of Paper~I. For every choice of the direction $\bn$ and the complex phase velocity $u$, one can find explicitly the four corresponding, generally complex values of $\kappa$. The true eigenmodes correspond to those having a real, positive value of $\kappa$. As we will see, this alternative formulation makes it much easier to understand the analytical properties of the dispersion relation.

\section{Lepton number conservation and transversality conditions}
\label{sec:lepton_number_transversality}

The dispersion relation we have derived so far is in some sense redundant. The reason is that we have not considered a fundamental constraint that the QKEs obey, namely the conservation of lepton number. 
From Eq.~\eqref{eq:EOM-linear}, it follows directly by summation that
\begin{equation}
    \partial_t \psi_0+i\bK\cdot\bpsi=0,
\end{equation}
where we have introduced the notation $\psi^\mu=(\psi_0,\bpsi)$.
From here it follows that for an eigenmode with frequency $\Omega$, the variable $\psi_0$ is constrained to be
\begin{equation}\label{eq:psi0_def}
    \psi_0=\frac{\bK\cdot\bpsi}{\Omega}
\end{equation}
and therefore not an independent degree of freedom.

The situation is somewhat analogous to the gauge invariance of the electrodynamical potentials. A certain gauge choice enforces a relation between the vector and scalar potential, so the latter is not an independent degree of freedom in an electromagnetic wave. The analogy is made yet clearer when we notice that Eq.~\eqref{eq:psi0_def} can be rewritten as a transversality condition
\begin{equation}
    \psi_\mu K^\mu=0,
\end{equation}
exactly as the electromagnetic four-potential is transverse to the wavevector. In this sense, it is useful to introduce a nomenclature based on this analogy, dubbing the three-dimensional vector $\bpsi$ the polarization vector of the flavor wave. We will also introduce the concept of transverse flavor waves, such that $\bK\cdot\bpsi=0$, and longitudinal ones, such that $\bpsi$ is aligned with $\bK$.

Transversality enforces a constraint on the dispersion relation itself, which is most easily obtained by breaking covariance; we note that from Eq.~\eqref{eq:EOM-linear} we can write
\begin{equation}
    \psi_\bv=\frac{G_\bv(\bK-\Omega \bv)\cdot\bpsi}{\Omega\,(\omega-\bk\cdot\bv)}.
\end{equation}
Upon multiplying by $\bv$ and integrating, we obtain a homogeneous relation for $\bpsi$ which admits a solution only if
\begin{equation}\label{eq:dispersion_relation_3d}
    \mathrm{det}\left[\delta_{ij}-\frac{K_i}{\Omega} I^{(1)}_j+I^{(2)}_{ij}\right]=0,
\end{equation}
where
\begin{equation}
    I^{(1)}_j=\int d^2\bv \frac{G_\bv v_j}{\omega-\bk\cdot\bv+i\epsilon}
    \quad\hbox{and}\quad
    I^{(2)}_{ij}=\int d^2\bv \frac{G_\bv v_i v_j}{\omega-\bk\cdot\bv+i\epsilon}.
\end{equation}
Here we have restored the $i\epsilon$ prescription, to be interpreted in the same sense as in Sec.~\ref{sec:dispersion_recap}. The tensor appearing in the determinant is itself not symmetric, due to the presence of the term $K_i I_j^{(1)}$. 

From their final forms, the equivalence between Eqs.~\eqref{eq:dispersion_relation_3d} and~\eqref{eq:dispersion_relation_4d} is not at all trivial to see; it is however guaranteed from the conservation law, so that all the solutions of the latter automatically also satisfy the former. This new dispersion relation may allow for some practical simplification, since it requires only a three-dimensional, rather than a four-dimensional, integral. On the other hand, it also comes with disadvantages, since it is not relativistically covariant, and it depends not only on the variable $\omega$ and $\bk$, but also explicitly on $\Omega$ and $\bK$. This makes it much more difficult to express results in terms of the complex phase velocity $u=\omega/\kappa$ introduced earlier. Nonetheless, we will show that the reduction in the dimensionality of the dispersion relation can still be of help in the solution of axisymmetric problems.

\section{Weak instabilities and resonance}
\label{sec:weak_instabilities_resonance}

The new perspective on the dispersion relation offered by the linear-response approach coincides with the conventional one for growing modes. However, it provides a new and important advantage; its analyticity means that even when we get to an infinitesimal growth rate, the function is well-behaved and can be differentiated. This property has a direct physical consequence. In our recent work~\cite{Fiorillo:2024qbl}, we have shown that most reasonably as soon as an angular distribution with a very small growth rate is formed, it will in some sense not evolve any further, remaining at the edge of instability. Thus, the regime of weak instabilities with very small growth rates deserves systematic investigation, which is most easily done using a dispersion relation that can be differentiated close to stability. 

In this section, we outline a general strategy to simplify a general dispersion relation of the form $\Phi(u,\kappa,\bn)=0$ when $\mathrm{Im}(u)\ll \mathrm{Re}(u)$. (We recall that $u=\omega/|\bk|$ is the complex phase velocity, $\kappa=|\bk|$ the modulus of the wavevector, and $\bn=\bk/|\bk|$ its direction.) The corresponding modes will be dubbed weakly unstable modes (or weakly damped modes if $\mathrm{Im}(u)<0$). For our case, $\Phi(u,\kappa,\bn)=\mathrm{det}\left[\kappa g^{\mu\nu}-\tilde{\chi}^{\mu\nu}(\kappa,\bn)\right]$.
For now, we do not deal with practical applications of this strategy; later, we will use it to obtain approximations for the growth rate of weakly unstable modes. Our dispersion relation is qualitatively different for $|\mathrm{Re}(u)|\leq1$ (subluminal modes) and $|\mathrm{Re}(u)|>1$ (superluminal modes). Hence, we keep the discussion separate for these two regimes.

\subsection{Weak superluminal instability}\label{sec:weak_superluminal}

In the superluminal regime, where $|\mathrm{Re}(u)|>1$, the denominators in $\tilde{\chi}^{\mu\nu}$ never vanish. Hence, if $u$ is real, the function $\Phi(u,\kappa,\bn)$ is purely real -- Landau's $i\epsilon$ prescription and the more standard dispersion relation coincide in this region. We seek solutions of the dispersion relation with a very small imaginary part $u=\uR+i \uI$; we will quantify in a moment how small. Since the dispersion relation is purely real, for any value of $\uR$, there certainly are solutions with vanishing imaginary part $\uI=0$ and a certain value of the wavevector $\overline{\kappa}$. Thus, $\Phi(\uR,\overline{\kappa},\bn)=0$. A weakly unstable solution must therefore be sought very close to one such solution; it will correspond to a value of the wavevector $\kappa=\overline{\kappa}+\delta\kappa$ and a very small imaginary part $\uI$, so we can expand the dispersion relation
\begin{equation}\label{eq:superluminal_dispersion_expanded}
    \left(-\frac{\partial^2 \Phi}{\partial\uR^2}\frac{\uI^2}{2}+\frac{\partial\Phi}{\partial \overline{\kappa}}\delta \kappa\right)+i \uI\left(\frac{\partial\Phi}{\partial\uR}-\frac{\uI^2}{6}\frac{\partial^3\Phi}{\partial\uR^3}\right)=0.
\end{equation}
We can now quantify how small $\uI$ must be. For the Taylor expansion, we require not only that $|\uI|\ll |\uR|$. Since the function $\Phi$ is only defined for $|\uR|>1$, and is singular close to this point, the more stringent condition $|\uI|\ll |\uR|-1$
must be satisfied.

The real and imaginary part of Eq.~\eqref{eq:superluminal_dispersion_expanded} must separately vanish, giving two conditions for $\uI$ and $\delta\kappa$. One obvious solution is $\uI=0$ and $\delta\kappa=0$, which is the stable one we started with. Another solution requires
\begin{equation}\label{eq:approx_superluminal}
    \uI^2={6\,\frac{\partial\Phi}{\partial\uR}}\bigg/  
    {\frac{\partial^3\Phi}{\partial \uR^3}}.
\end{equation}
For a generic function $\Phi$, the order of magnitude of the right-hand side of this equation is $\uR^2$, thus showing that the imaginary part is typically comparable with the real part and violating our original assumption. Thus, for $\uR$ sufficiently far from $|\uR|=1$, weakly unstable modes can only exist close to the special point $\uR$ for which $\partial\Phi/\partial\uR=0$. At such a point, unless special fine tuned cases are considered $\partial^3\Phi/\partial\uR^3$ does not vanish. Thus, this point is the edge of the region where unstable modes exist. On one side, the right-hand side of Eq.~\eqref{eq:approx_superluminal} is negative and so no unstable modes exist, while on the other side it becomes positive, so a pair of complex conjugate modes appear.

Our conclusion is that outside of the luminal sphere, weak instabilities, with $|\uI|\ll |\uR|-1$, can only exist close to this special point. Still, for any given angular distribution some unstable superluminal modes must exist, given our proof in Ref.~\cite{Fiorillo:2024qbl} that close to a crossing line on the luminal sphere, it is certain that unstable modes exist. If an angular distribution has a very shallow angular crossing, there should not be modes with a growth rate of order $1$ (because of our choice $\mu=1$ this means growth rates much smaller than the typical ``fast'' ones), since by a small deformation the distribution can be distorted into one without an angular crossing. Therefore, for such a distribution, the unstable superluminal modes must necessarily terminate very close to the luminal sphere; this is the only possibility to have $|\uI|\sim |\uR|-1$ and yet have very small growth rates. For such modes, the Taylor expansion we have performed so far is not valid; we might call these near-luminal instabilities, which can be treated by methods analogous to those developed in Sec.~5 of our companion paper~\cite{Fiorillo:2024bzm}. We will verify in an explicit example later that indeed a distribution with a very shallow angular crossing possesses unstable superluminal modes only very close to the luminal sphere, as shown by our reasoning above.

Finally, one might wonder whether the system might admit instabilities that are not weak, in the sense that $\mathrm{Im}(u)$ is not much smaller than $\mathrm{Re}(u)$, but nevertheless with a growth rate $\mathrm{Im}(\omega)=\mathrm{Im}(u)\,\kappa \ll 1$. One easily sees that this can happen if $\kappa\ll 1$. Thus, this case is most easily treated by looking at the original version of the dispersion relation in Eq.~\eqref{eq:dispersion_relation_4d} in the limit $\kappa\to 0$ and $\Omega$ finite. For $\kappa=0$ the dispersion relation can be written as
\begin{equation}
    \mathrm{det}\left[\Omega g^{\mu\nu}-\int d^2\bv G_\bv v^\mu v^\nu\right]=0.
\end{equation}
In this range of wavevectors, the dispersion relation generically depends on the angle-integrated properties of the neutrino distribution. This should not come as a surprise, since these modes are strongly superluminal -- as $\kappa\to 0$ the phase velocity of the wave becomes infinite -- so that the modes grow from the non-resonant interaction with the entire angular distribution, rather than from the resonant interaction with specific neutrino modes. Generally speaking, for superluminal modes the often-used approximation of a few discrete neutrino beams is much more appropriate than for subluminal modes, given the non-resonant nature of the wave-particle interaction.

The eigenfrequencies in this limit are the eigenvalues of the tensor $\int d^2\bv G_\bv v^\mu v^\nu$. This is a symmetric tensor in a pseudo-Euclidean metric, and as we will see later, it may admit a pair of complex conjugate solutions. However, the imaginary part of these solutions will generally be of the order of 1, and therefore not small, unless the angular distribution is fine-tuned. So, it appears that eigenmodes with a small growth rate generally must be weak instabilities $\mathrm{Im}(u)\ll 1$, except for fine-tuned angular distributions.

To summarize our findings in this section: weakly unstable superluminal modes, in the sense that $|\mathrm{Im}(u)|\ll |\mathrm{Re}(u)|$, can only exist close to the edge of instability, defined by the condition $\partial \Phi/\partial u=0$. Close to the luminal sphere, defined by  $|\mathrm{Re}(u)|=1$, there can be modes with $\mathrm{Im}(u)\sim |\mathrm{Re}(u)|-1\ll 1$ with a very small growth rate. Finally, strongly superluminal modes with $|\mathrm{Re}(u)|\gg 1$, though not weakly unstable, can still have a small growth rate if they correspond to very small values of $\kappa$; the corresponding eigenfrequencies are most easily found as the eigenvalues of the tensor $\int d^2\bv G_\bv v^\mu v^\nu$, so their growth rate is small only if this tensor has a pair of eigenvalues with a small imaginary part.

\subsection{Weak subluminal instability}\label{sec:weak_subluminal}

For $|\mathrm{Re}(u)|\leq 1$, the nature of the weakly unstable modes is entirely different, because the function $\Phi(u,\kappa,\bn)$ has a non-vanishing imaginary part even for purely real $u$. This imaginary part arises from the Landau prescription of integration ($i\epsilon$ in the denominator) introduced in our companion paper~\cite{Fiorillo:2024bzm}; it is the price for a dispersion relation that can be differentiated. Physically it reflects Landau damping of subluminal modes and also implies that solutions no longer appear in complex conjugate pairs. Thus, it is no longer certain that for a real $u=\uR$ there are solutions of the dispersion relation with a real, positive value of $\kappa$. Nevertheless, we can still seek solutions $u=\uR+i\uI$ with $|\uI|\ll 1$ (this condition suffices here, while in the superluminal case the more lenient condition $|\uI|\ll |\uR|$ was sufficient) and $|\uI|\ll 1-|\uR|$. Expanding to first order in $\uI$ we find
\begin{equation}\label{eq:dispersion_subluminal}
    \mathrm{Re}(\Phi)-\frac{\partial\mathrm{Im}\Phi}{\partial\uR}\uI +i \left(\mathrm{Im}(\Phi)+\uI\frac{\partial\mathrm{Re}(\Phi)}{\partial\uR}\right)=0,
\end{equation}
where the function $\Phi$ is always evaluated for real $u=\uR$. Since the imaginary part must separately vanish, we find
\begin{equation}\label{eq:weak_subluminal_growth}
    \uI=-{\mathrm{Im}(\Phi)}\bigg/{\frac{\partial\mathrm{Re}(\Phi)}{\partial\uR}}.
\end{equation}
The order-of-magnitude of this quantity is $\uI\sim\uR \mathrm{Im}(\Phi)/\mathrm{Re}(\Phi)$. Assuming $\uR\sim 1$, we see that weakly unstable modes are naturally found if the imaginary part of $\Phi$, evaluated on the real axis, is much smaller than the real part. This condition directly matches the intuitive nature of instability we have introduced in Paper~I, where we showed that as soon as an angular distribution develops a crossing, the waves resonant with the neutrinos on the weak side of the crossing are amplified. In order for the growth rate to be small, there must be a small amount of lepton number on the weak side of the crossing, so that the distribution $G_{\bv}$ in that region must be small. It follows that the imaginary part $\mathrm{Im}(\Phi)$ must be small; with the arguments presented here, we convert this intuitive picture into a formal criterion.

If this condition is satisfied, the vanishing of the real part implies $\mathrm{Re}(\Phi)=0$. In other words, for weakly unstable modes, the real part of the phase velocity $\uR$ is determined by the real part of the dispersion relation, while the imaginary part $\uI$ is proportional to the imaginary part of the function $\mathrm{Im}(\Phi)$ evaluated on the real axis of $u$. However, Eq.~\eqref{eq:dispersion_subluminal} is more general and does not require $\mathrm{Im}(\Phi)\ll \mathrm{Re}(\Phi)$; for a fixed $\uR$, the vanishing of the real and imaginary parts gives two conditions for $\kappa$ and $\uI$. Since $\Phi$ is a polynomial function of $\kappa$, finding the solutions of these two equations does not require to solve a transcendental equation, but rather a simple polynomial one, considerably simplifying the search for unstable modes. If $\mathrm{Im}(\Phi)$ is only marginally smaller than the real part, then we may treat it as a small perturbation. To lowest order, the value of the wavenumber associated with a given phase velocity can be written as $\kappa=\overline{\kappa}+\delta\kappa$, with $\overline{\kappa}$ determined by the condition $\mathrm{Re}\left[\Phi(\uR,\overline{\kappa})\right]=0$. At the next order, we can find the displacement $\delta\kappa$
\begin{equation}
    \frac{\partial\mathrm{Re}(\Phi)}{\partial\kappa}\delta\kappa-\frac{\partial\mathrm{Im}(\Phi)}{\partial\uR}\uI=0,
\end{equation}
and replacing for $\uI$ the expression in Eq.~\eqref{eq:weak_subluminal_growth} we find
\begin{equation}\label{eq:approx_delta_kappa}
    \delta\kappa=-{\mathrm{Im}(\Phi) \frac{\partial\mathrm{Im}(\Phi)}{\partial\uR}}\bigg/{\frac{\partial\mathrm{Re}(\Phi)}{\partial\uR}\frac{\partial\mathrm{Re}(\Phi)}{\partial\kappa}},
\end{equation}
where all the functions on the right-hand side are evaluated at $\kappa=\overline{\kappa}$. Equations~\eqref{eq:weak_subluminal_growth} and~\eqref{eq:approx_delta_kappa} give a direct algorithmic procedure to find the wavevector and growth rate of weakly unstable subluminal modes with phase velocity $\uR$.

From the definition of $\Phi$, we easily see that the imaginary part of $\Phi$ on the real axis of $u$ arises from the imaginary part of
\begin{eqnarray}
    \tilde{\chi}^{\mu\nu}&=&\int d^2\bv\,\frac{G_\bv v^\mu v^\nu}{\uR-\bn\cdot\bv+i\epsilon}
    \nonumber\\[1.5ex]
    &=&\fint d^2\bv\,\frac{G_\bv v^\mu v^\nu}{\uR-\bn\cdot\bv}-i\pi \int d^2\bv\,G_\bv v^\mu v^\nu \delta(\uR-\bn\cdot\bv).
\end{eqnarray}
The imaginary part receives contributions only from resonant neutrinos satisfying the condition $\uR=\bn\cdot\bv$. These neutrinos have a velocity component in the direction of the wave equal to the phase velocity of the wave itself. So the growth rate for weak subluminal instabilities is determined by the number of neutrinos resonantly moving with the wave. This is perhaps our central result and it directly descends from the intuitive explanation of fast instabilities introduced in Paper~I. They correspond to flavor waves feeding on the Cherenkov emission of individual, resonant neutrinos; therefore, their growth rate must be proportional to the amount of resonant neutrinos. 

\subsection{Angular crossings and instabilities}\label{sec:angular_crossings}

The concept of instability as driven by a resonant interaction with individual neutrinos allowed us, in Paper~I~\cite{Fiorillo:2024bzm}, to prove that an angular crossing always leads to an instability. Here we briefly recap the physical argument. By angular crossing we mean a line on the unit sphere defined by the directions $\bv$ through which the distribution $G_\bv$ changes sign. The gist of the argument is that, in the presence of an angular crossing, flavor waves resonant with neutrinos on opposite sides of the crossing feel an opposite sign for the resonant interaction. Therefore, modes resonant with neutrinos on one of the two sides necessarily must be unstable. In other words, the crossing line is an edge in the space of directions $\bv$, bounding a region of ``flipped neutrinos;'' there will be some modes pointing in the same direction as the flipped neutrinos which are unstable. The notion of a flipped-neutrino region becomes particularly clear in the case of weak instabilities, which appear when an originally stable distribution, with no angular crossing, is slowly deformed into an unstable one, with an angular crossing. The small deformation leads to a region where $G_\bv$ is small and with opposite sign to the undeformed value: this is the flipped region. 

Geometrically, the correspondence between the modes and the resonant neutrinos is established in the simplest way by introducing the phase velocity vector of the wave $\bu=\mathrm{Re}(u)\,\bn$. Modes for which $\mathrm{Re}(u)=1$ are luminal, moving with the speed of light; we will call the surface of these modes the luminal sphere. The special status of luminal modes descends from the fact that they are only resonant with neutrinos exactly collinear with their direction. In the space of the phase velocity vector $\bu$, we can identify regions for which there are unstable modes, i.e., regions of instability. Our argument above, proven formally by algebraic means by Morinaga~\cite{Morinaga:2021vmc} and constructively in our Paper~I \cite{Fiorillo:2024bzm}, shows that the crossing lines on the luminal sphere must always correspond to an edge for the region of instability, and that luminal modes with $\bn$ pointing in the flipped region must be unstable. We will later make use of this geometrical argument to discuss the special case of axisymmetric distributions, and verify it explicitly for an example distribution.

\subsection{Boundaries of the region of instability}\label{sec:boundaries_general}

Regardless of whether the angular distribution leads only to weakly unstable modes -- as expected for evolution along the edge of instability~\cite{Fiorillo:2024qbl} -- for any angular distribution there are values of $\bu$, or equivalently of $\bk$, for which the growth rate is small. These correspond to the boundaries of the region of instability. With the approximate expansions that we have provided for weakly unstable modes, we can now provide explicit criteria to identify these boundaries.  We assume a fixed direction $\bn$, and seek the values of $\mathrm{Re}(u)$, or equivalently of $\kappa$, corresponding to the edge of the instability region.

For superluminal instabilities, as we have already discussed, the boundary of the instability region can be found by the condition ${\partial \Phi}/{\partial\uR}=0$. Here the derivative is computed for a purely real value of the phase velocity $u$, and the function $\Phi$ is naturally real in the superluminal region. Let us call $\overline{\kappa}$ the value of the wavevector for which ${\partial\Phi}/{\partial\uR}=0$; close to this point, we may expand this function to first order in $\delta\kappa=\kappa-\overline{\kappa}$. From Eq.~\eqref{eq:approx_superluminal}, it follows that asymptotically close to the boundary of the instability region, the growth rate scales as $\uI\propto \sqrt{\delta\kappa}$; superluminal modes branch with a characteristic square-root behavior. This should not come as a surprise, since in the superluminal region the dispersion relation is purely real and modes must appear always in pairs of complex conjugates, enforcing the square-root behavior.

For subluminal instabilities, we find from Eq.~\eqref{eq:weak_subluminal_growth} that the boundary of the instability region corresponds to $\mathrm{Im}(\Phi)$ changing sign. Thus, we can look for this boundary through the condition $\mathrm{Im}(\Phi)=0$, again evaluated for real values of the phase velocity $u$, with $\mathrm{Im}(\Phi)$ changing sign in passing through this point. In this case, close to the value $\overline{\kappa}$ for which $\mathrm{Im}(\Phi)=0$, we can again perform a Taylor expansion to find that $\uI\propto \delta\kappa$.

Finally, on the luminal sphere, where $\mathrm{Re}(u)=1$, the boundary of the instability region is constituted by the crossing lines on which $G_{\bu}$ vanishes as discussed earlier.

These criteria allow us to identify the regions of instability for a generic distribution, without necessarily having to find the growth rate itself. Later, we will apply them to the special case of axisymmetric, single-crossed distributions to obtain explicit expressions for the instability regions of modes directed along the axis of symmetry.

\section{Axisymmetric distributions}\label{sec:axisymmetric_distributions}

Most works on collective neutrino conversions have considered either systems with axial symmetry or, even more idealized, one-dimensional cases. Therefore, it may be useful to restrict ourselves to axial symmetry and discuss how our general results for weak instabilities apply in this more symmetric case.

\subsection{General properties and dispersion relation}

By definition, an axisymmetric distribution $G_{\bv}$ depends only on the polar angle with respect to a certain axis of symmetry $\hat{\bz}$, so we can write $G_{\bv}=G_v$, where $v=\bv\cdot \hat{\bz}$. On the other hand, it is sometimes forgotten that even an axisymmetric distribution can be unstable to axial-breaking perturbations that will grow and break spontaneously the original symmetry. Mathematically, even though the distribution $G_{v}$ is axisymmetric, generally there will be unstable modes for which either the wavevector $\bk$ or the polarization vector $\bpsi$ or both are not aligned with the axis of symmetry. In contrast, many previous works have focused on the purely one-dimensional case where everything depends only on the coordinate $z$ along the axis of symmetry -- so considering only modes with $\bk\parallel \bz$ -- and the perturbed field $\psi_\bv$ also depends only on $v$ -- so that $\bpsi\parallel\bz$. While there is nothing wrong with considering such symmetric solutions, a few issues deserve clarification. 

First, a crossing of an axisymmetric $G_v$ does not guarantee unstable solutions with the same symmetry. On the contrary, Morinaga's proof~\cite{Morinaga:2021vmc} and also our intuitive argument in Sec.~\ref{sec:angular_crossings} only guarantee unstable modes with wavevectors pointing toward the crossing line on the unit sphere and therefore away from the axis of symmetry. Stable modes are on one side of the crossing line, unstable ones on the other, encompassing the region of flipped neutrinos. If the flipped region does not include one of the poles, along the positive or negative $z$ axis, then unstable modes directed along $z$ are not guaranteed. Typically, however, distributions with a single crossing were assumed. In this case, the positive and negative $z$ axes intersect the luminal sphere at $v=\pm 1$, and the two values $G_v$ at these points have opposite signs. This means that one of the two must be in the flipped region. So for single-crossed axisymmetric distributions (or more generally axisymmetric distributions with an odd number of crossings) there are always unstable modes directed along the axis of symmetry. An explicit criterion for double-crossed distributions without instabilities along the axis of symmetry was given previously \cite{Capozzi:2019lso}, which is a special form of the general criterion to find the region of instability we introduced in Sec.~\ref{sec:boundaries_general} and that we later specialize to axisymmetric distributions.

Second, even though for single-crossed distributions there always exist axisymmetric unstable modes, there are always also axially-breaking ones. It is possible that the average of the resulting solution in the nonlinear regime over sufficiently large regions will be the same, regardless of whether these axially-breaking modes are included or not; after all, quasi-linear theory for a limited number of beams shows that the system anyway settles into a space-averaged configuration which is linearly stable~\cite{Fiorillo:2024qbl}. Symmetry arguments may be enough to contend that the space-averaged density matrix, in an axisymmetric system, must also be axisymmetric, so perhaps the axially-breaking modes would indeed not affect the space-averaged results, but this is a speculation. For the purpose of numerical solutions, there is no justification to leave out axially-breaking modes, which are unstable with comparable growth rates as the axially symmetric ones.

For axisymmetric systems, some properties of the eigenmodes of the flavor susceptibility $\chi^{\mu\nu}$ (or equivalently of $\tilde{\chi}^{\mu\nu}$) can be proven in general. Let the wavevector $\bk$ lie in the $x$--$z$ plane, so we may write it as $\bk=\kappa\,(\cos\alpha\,\hat{\bf x}+\sin\alpha\,\hat{\bz})$. Symmetry implies $\chi^{0y}=\chi^{xy}=\chi^{zy}=0$, so for any direction of $\bk$ we can always find one eigenmode $\bpsi$ directed along $y$. Such an eigenmode is transverse to the plane containing $\bk$ and the axis of symmetry, and we will dub it the transverse mode T. Its dispersion relation is given by 
\begin{equation}\label{eq:dispersion_relation_t2_mode}
    \tilde{\chi}^{yy}=\int dvd\phi\,\frac{G_v (1-v^2) \sin^2\phi}{u-v\cos\alpha-\sqrt{1-v^2}\sin\alpha \cos\phi+i\epsilon}=-\kappa.
\end{equation}
As usual, $i\epsilon$ in the denominator serves as a reminder of the prescription for integration. For luminal modes with no imaginary part ($u=1$), resonant neutrinos must move collinear with the wave, and therefore $v=\cos\alpha$ and $\phi=0$; for such neutrinos, the numerator in the integrand vanishes, implying that $\tilde{\chi}^{yy}$ is purely real. This proves that transverse modes are stable on every point of the luminal sphere, not just at the crossing line.

The remaining components of the flavor susceptibility form a $3\times 3$ matrix, and cannot be separated in a general form. On the other hand, it can be further simplified for modes directed along the axis of symmetry, with $\alpha=0$ or $\pi$. In this case, the mode polarized with $\bpsi\parallel\hat{\bx}$ is degenerate with the transverse mode T and has the same dispersion relation; we call this additional mode T1. We emphasize that two transverse modes exist only for $\bk$ parallel to the symmetry axis. Their dispersion relation immediately follows from Eq.~\eqref{eq:dispersion_relation_t2_mode}. Rather than writing it in terms of the module $\kappa$, with $\alpha=0$ or $\pi$, we write it in terms of $k_z$, which can be either positive or negative, and $u_z=\omega/k_z$, so that
\begin{equation}
    \int_{-1}^{+1}\!dv\,\frac{g_v (1-v^2)}{u-v+i\epsilon s_z}+2k_z=0,
\end{equation}
where $s_z=\mathrm{sign}(k_z)$, so $u_z=s_z u$, and we have introduced $g_v=2\pi G_v$ as the neutrino distribution already integrated over the azimuthal angle. Introducing the notation
\begin{equation}
    I_n=\int_{-1}^{+1}\!dv\, \frac{g_v v^n}{u_z-v+i\epsilon s_z},
\end{equation}
where the integrals are functions of the phase velocity $u_z$, we can compactly write
\begin{equation}\label{eq:transverse_dispersion_relation}
    I_0-I_2+2k_z=0
\end{equation}
for the dispersion relation of these modes.

The remaining components of the flavor susceptibility form a $2\times 2$ matrix, corresponding to purely longitudinal modes, namely the ones that are usually considered in the literature. Their dispersion relation can be obtained by diagonalizing the corresponding matrix; however, it is easier to find it using the transversality constraint introduced in Sec.~\ref{sec:lepton_number_transversality}. The vector $\bpsi=\psi_z \hat{\bz}$ is directed along the $z$ axis. We start from the zeroth component of the full dispersion relation in Eq.~\eqref{eq:wrong_chi}, after changing in the denominator $k\cdot v\to k\cdot v+i\epsilon$ by the usual prescription,
\begin{equation}
    \int_{-1}^{+1}\!dv\, \frac{g_v(\psi^0-v\psi_z)}{u_z-v+i\epsilon s_z}=k_z \psi^0.
\end{equation}
From the transversality condition in Eq.~\eqref{eq:psi0_def}, for modes along the axis of symmetry, we obtain $\psi_z=\psi_0 \Omega/K_z=\psi_0(\omega-G_0)/(k_z-G_1)$, implying
the dispersion relation
\begin{equation}
    \int_{-1}^{+1}\!dv\, \frac{g_v}{u_z-v+i\epsilon s_z}\left(1-\frac{v(\omega-G_0)}{k_z-G_1}\right)=k_z.
\end{equation}
Since $\omega=u_zk_z$, this expression can be brought into the form
\begin{equation}\label{eq:dispersion_relation_axial_kappa}
    k_z^2-k_z(I_0-I_2)+G_1 I_0-G_0 I_1=0.
\end{equation}
For each value of $u_z$, Eq.~\eqref{eq:dispersion_relation_axial_kappa} leads to two values of $k_z$ corresponding to the two longitudinal modes. For temporally unstable modes, $k_z$ must be real, so the dispersion relation is again only implicit, because it does not provide an explicit dependence of $u_z$ on $k_z$.

\subsection{Boundaries of the region of instability}\label{sec:boundaries_axial}

With these simplified forms for the dispersion relation of modes directed along the axis of symmetry, we can identify the boundaries of the region of instability, i.e., the values of $k_z$ marking the interval in which unstable modes exist. We focus on single-crossed distributions, because for multiple-crossed ones, unstable modes along the axis of symmetry may not even exist as discussed earlier.

We begin with superluminal modes, i.e., those with $|\mathrm{Re}(u_z)|>1$. In this case, as we have discussed in Sec.~\ref{sec:weak_superluminal}, unstable modes branch from a value of $k_z$ such that both $\Phi(u_z,k_z)=0$ and $\partial\Phi(u_z,k_z)/\partial u_z=0$, where $u_z$ is now interpreted as a real variable. For the transverse modes, this corresponds to
\begin{equation}\label{eq:twoconditions}
    I_0-I_2+2k_z=0
    \quad\hbox{and}\quad
    F_0-F_2=0,
\end{equation}
where
\begin{equation}
    F_n=\int dv\,\frac{g_v v^n}{(u_z-v)^2}.
\end{equation}
Notice that no $i\epsilon$ prescription is needed for superluminal phase velocities. Equation~\eqref{eq:twoconditions} specifies two conditions for the two unknowns $u_z$ and $k_z$, allowing us to determine explicitly the threshold value of $k_z$ for the appearance of unstable modes. For a single-crossed distribution, they always admit a solution. To see this, we note that for $u_z\to \pm\infty$ the function $F_0-F_2$ has the same sign, either positive or negative, since $F_0-F_2\simeq (g_0-g_2)/u_z^2$. On the other hand, for $u_z=\pm 1\pm \delta u_z$, the function $F_0-F_2\sim 2 g(\pm 1) \log(2/\delta u_z)$ is logarithmically divergent. Since $g(\pm 1)$ has opposite signs for a single-crossed distribution, it follows that either in the range $]{-}\infty,{-}1[$ or in the range $]{+}1,{+}\infty[$ the function has to change sign, and therefore a zero of $F_0-F_2$ must exist. The corresponding value of $k_z$ is determined by the first equation $k_z=(I_2-I_0)/2$, so there must be a superluminal boundary to the region of instability. This is in fact easier to see by a different argument; since the instability range has a single subluminal boundary, as we show below, it must necessarily end at a superluminal boundary.

Similarly, for the longitudinal modes, we can determine the threshold values of $k_z$ and $u_z$, again interpreted as a real variable, for which unstable superluminal modes appear from the joint conditions
\begin{equation}
    k_z^2-k_z(I_0-I_2)+G_1 I_0-G_0 I_1=0
    \quad\hbox{and}\quad
    -k_z(F_0-F_2)+G_1F_0-G_0F_1=0.
\end{equation}
In this case, there is not necessarily a superluminal boundary to the region of instability, so this equation does not necessarily admit a real solution for $k_z$. However, for angular distributions that are very close to stability such a boundary does exist. The reason is that there are two modes becoming unstable for subluminal phase velocity, as we show below; the threshold value of phase velocity is $u_z=v_{\rm cr}$. The growth rates for these modes must then close either for superluminal phase velocities (it cannot have another zero in the subluminal range, see below) or they must remain positive up to $|u_z|\to \infty$. In the latter case, the range of values of $k_z$ containing unstable modes would be delimited by the two subluminal boundaries. In this case, however, as we discussed in Sec.~\ref{sec:weak_superluminal}, one does not expect very small growth rates for $|u_z|\gg 1$. Therefore, for a weakly unstable angular distribution, two superluminal boundaries are expected as the closure of the corresponding two modes becoming unstable in the subluminal range.

We can now discuss the threshold values of $k_z$ and $u_z$ for the existence of unstable subluminal modes. As we have discussed in Sec.~\ref{sec:weak_subluminal}, these correspond to the points in which the function $\mathrm{Im}\left[\Phi(u_z,k_z)\right]$ changes sign for a real value of $|u|<1$. Here we will use repeatedly the standard property that
\begin{equation}
    \mathrm{Im}\int dv\frac{f(v)}{u_z-v+i\epsilon s_z}=-\pi f(u_z) s_z.
\end{equation}
For the transverse modes, the imaginary part of $\Phi(u_z,k_z)$ vanishes either for $u_z=\pm 1$ and for $u_z=v_{\rm cr}$, where $v_{\rm cr}$ is the crossing velocity at which $g_v$ changes sign. The vanishing for $u_z=\pm 1$ is at the edge of the subluminal region and therefore does not correspond to the appearance of unstable modes. Thus, the boundary for the appearance of unstable subluminal modes must correspond to $u_z=v_{\rm cr}$. The corresponding value of $k_z$ is then easily found from
 \begin{equation}\label{eq:threshold_transverse}
     k_z=\frac{I_2(v_{\rm cr})-I_0(v_{\rm cr})}{2},
 \end{equation}
 where we now specify the dependence on $u_z$ evaluated at $u_z=v_{\rm cr}$; of course this value of $k_z$ is real, as it should.

 For longitudinal modes, the dispersion relation in Eq.~\eqref{eq:dispersion_relation_axial_kappa} is quadratic and therefore slightly more involved. The imaginary part of the dispersion relation vanishes for
 \begin{equation}
     g(u_z)\left[-k_z(1-u_z^2)+G_1-G_0 u_z\right]=0,
 \end{equation}
 and it changes sign either for $u_z=v_{\rm cr}$ or for
 \begin{equation}
     u_\pm=\frac{G_0\pm\sqrt{G_0^2-4G_1 k_z+4k_z^2}}{2k_z}.
 \end{equation}
 For the case of longitudinal modes, the fact that an instability must appear in correspondence to one of these critical points was noted in Ref.~\cite{Capozzi:2019lso}, with the empirical observation that usually only the critical point corresponding to the crossing $u_z=v_{\rm cr}$ is the true boundary of unstable modes. Actually, we can prove explicitly that this is always the case. The dispersion relation evaluated at each of the two points $\Phi(u_\pm(\kappa),\kappa)$ is equal to

 \begin{equation}
     \Phi(u_\pm(k_z),k_z)=k_z^2\left[1+\frac{u_\mp(k_z) G_0-G_1}{k_z}\right].
 \end{equation}
By explicit substitution, one easily verifies that for both solutions $u_\pm(k_z)$ this quantity is always positive for any value of $k_z$, and therefore can never correspond to a threshold eigenmode of the dispersion relation. There is one exception, for $k_z=G_1$, corresponding to the homogeneous eigenmode with $K_z=0$, for which the solution with $u_+=G_0/G_1$ (or $u_-=G_0/G_1$ if $G_0<0$) leads to $\Phi(u_+(G_1),G_1)=0$. However, this solution cannot correspond to a boundary of the region of instability, as one can easily show by noting that close to it $\Phi(u_+(k_z),k_z)>0$ for any value of $k_z$, implying that at $k_z=G_1$ one has
\begin{equation}
    \frac{\partial \Phi}{\partial u}\frac{du_+(k_z)}{dk_z}+\frac{\partial\Phi}{\partial k_z}=0.
\end{equation}
Thus, if we move from $k_z=G_1$ to $k_z=G_1+\delta k_z$, the corresponding solution changes by 
\begin{equation}
    \delta u=-\delta k_z \frac{\frac{\partial \Phi}{\partial k_z}}{\frac{\partial\Phi}{\partial u}}=\frac{du_+(k_z)}{dk_z}\delta k_z,
\end{equation}
and therefore it develops no imaginary part on either side. Later, we will show that the same result follows from an application of Nyquist's criterion. Thus, neither of the two solutions $u=u_\pm(k_z)$ can correspond to a boundary for the interval of unstable wavenumbers.

The only remaining possibility is $u=v_{\rm cr}$. Solving the quadratic equation~\eqref{eq:dispersion_relation_axial_kappa} for the integrals $I_n$ evaluated at $u=v_{\rm cr}$, we find immediately the explicit expressions for the two threshold values $k_{\pm}(v_{\rm cr})$ for the two longitudinal, subluminal modes. These coincide with the threshold values obtained by similar methods in Ref.~\cite{Capozzi:2019lso}. Their explicit values are
\begin{equation}\label{eq:threshold_longitudinal}
    k_\pm=\frac{I_0-I_2\pm\sqrt{(I_0-I_2)^2+4(G_0 I_1- G_1 I_0)}}{2},
\end{equation}
which can be simplified noting the relation
\begin{equation}
    G_n=v_{\rm cr}I_n-I_{n+1},
\end{equation}
leading to
\begin{equation}
    k_\pm=\frac{I_0-I_2\pm\sqrt{(I_0-2I_1+I_2)(I_0+2I_1+I_2)}}{2}.
\end{equation}
The factor under square root is always positive, since
\begin{equation}
    I_0\pm 2I_1+I_2=\int dv\frac{g(v)}{v_{\rm cr}-v}(1\pm v)^2
\end{equation}
has an integrand function that is always positive (assuming $g_v>0$ for $v<v_{\rm cr}$; one can always redefine the sign of $g_v$ so that this condition is satisfied). Therefore, for single-crossed distributions, we find that there are always unstable axially symmetric eigenmodes, in agreement with our geometric argument given earlier.

The existence of these threshold values can be understood also on the basis of a method more conventional in plasma physics, namely the Nyquist criterion~\cite{nyquist1932regeneration}. We have previously introduced this method for the special case of homogeneous instabilities, i.e., with $K_z=0$ or $k_z=G_1$ \cite{Fiorillo:2023hlk}. The properties of such instabilities are made special by the existence of a host of conserved quantities, first proposed in Ref.~\cite{Johns:2019izj} and later explicitly identified in Refs.~\cite{Fiorillo:2023mze, Fiorillo:2023hlk}. Here we briefly discuss how Nyquist's criterion relates to the previous argument on the boundaries of the instability region. We focus on the case of longitudinal modes only; the reasoning applies in the same way to the transverse modes as well.

We are seeking the number of zeroes of the function $\Phi(u_z,k_z)$ in the upper half-plane of the complex variable $u_z$, namely the number of unstable modes.
Nyquist's criterion ensures that this number is equal to the number of times the function $\Phi(u_z,k_z)$, in the complex plane, encircles $0$ as $u$ progresses along the real axis from $-\infty$ to $+\infty$ for a fixed value of~$k_z$. We call this the winding number. We will now show that this number can easily be found by a geometrical argument. For definiteness, we assume again $G_0>0$.

For $u_z\to -\infty$ and $u_z\to+\infty$, the function $\Phi(u_z)$ in the complex plane starts and ends at $\Phi(u_z)\to k_z^2$. As $u$ increases from $-\infty$, the function acquires an imaginary part as soon as $u_z>-1$ and describes a trajectory in the complex plane. The key insight is that the function $\Phi(u,\kappa)$ intersects the real axis precisely at the critical points we have identified before where $\mathrm{Im}\left[\Phi(u_z,k_z)\right]=0$, namely $u_z=u_\pm, v_{\rm cr}$. The first two solutions always lead to intersections on the positive side of the real axis; since the function $\Phi(u_z,k_z)$ started on the same positive side for $u_z\to \pm \infty$, these intersections do not allow to encircle the point $\Phi=0$. The only exception is again the case $k_z=G_1$, for which $\Phi[u_+(G_1),G_1]=0$; however, in this case a small change in the value of $k_z\to k_z+\delta k_z$ always moves the intersection with the real axis on the positive side, and therefore this value cannot be the boundary of a region of instability. Thus, the \textit{necessary} condition to possess an unstable mode is that for the other critical point $u_z=v_{\rm cr}$ the function $\Phi(v_{\rm cr},k_z)<0$. The threshold value of $k_z$ for which unstable modes start to exist is then given by the condition $\Phi(v_{\rm cr},k_z)=0$, which is exactly the condition we found before. However, by Nyquist's criterion, we also identify directly on which side of the boundary there is a region of instability, namely the values of $k_z$ for which $\Phi(v_{\rm cr},k_z)<0$. We should specify, though, that by this form of Nyquist's criterion we are only able to identify the subluminal thresholds for the region of instability. There might be superluminal thresholds, which cannot easily be identified by this criterion, because for superluminal phase velocities the function $\Phi(u_z,k_z)$ is real, and therefore it might cross $0$ exactly on the real axis. In this situation, the winding number is ill-defined, and Nyquist's criterion fails. We will investigate this particular question in an upcoming work~\cite{NyquistPaper}.

We should also comment on the relation of the new criterion introduced here with our previous criterion for homogeneous instabilities. Given that we have identified two subluminal thresholds $k_\pm$ for the range of unstable wavenumbers, one might conclude that the homogeneous mode $K_z=0$, i.e.\ $k_z=G_1$, is unstable if $k_-<G_1<k_+$. By replacing the values of $k_\pm$, this would lead to the condition
\begin{equation}\label{eq:naive_condition}
    I_1(v_{\rm cr})(G_1 v_{\rm cr}-G_0)<0.
\end{equation}
Here the integral $I_1$ is evaluated at $u=v_{\rm cr}$. However, this condition turns out to be not sufficient for an instability to appear. The reason was already identified various times above; for $k_z=G_1$, the function $\Phi(u_z,G_1)$ has always a zero at $u_z=G_0/G_1$, which coincides with $u_+(G_1)$. In other words, the dispersion relation \textit{always} admits a homogeneous eigenmode with $\omega=G_0$, or $\Omega=0$. Indeed, by replacing $k_z=G_1$ we find
\begin{equation}
    \Phi(u_z,k_z)=I_1(u_z)(G_1 u_z-G_0).
\end{equation}
The condition in Eq.~\eqref{eq:naive_condition} implies that $\Phi(u_z,k_z)$ has a zero in the upper half-plane (including the real axis), but this zero may coincide with the trivial zero $u_z=G_0/G_1$. Therefore, in order to have an instability, one should require the more stringent condition that only $I_1(u)=0$ has a zero in the upper half-plane. This coincides with the condition we introduced in Ref.~\cite{Fiorillo:2023hlk}, which led us to the more stringent criteria
\begin{equation}
    \frac{G_1}{G_0 v_{\rm cr}}<0
    \quad\hbox{and}\quad
    \frac{I_1(v_{\rm cr})}{G_0}>0.
\end{equation}
Notice that these can be rewritten as
\begin{equation}
    I_1(v_{\rm cr})v_{\rm cr}G_1<0
    \quad\hbox{and}\quad
    I_1(v_{\rm cr})G_0>0.
\end{equation}
Obviously together they imply Eq.~\eqref{eq:naive_condition}, but they are more stringent. 

\subsection{Growth rate of weak subluminal instabilities}
\label{sec:growth_axi_weak_subluminal}

We now apply the general strategy of approximation introduced in Sec.~\ref{sec:weak_subluminal} to obtain explicit expressions for the growth rate of weak subluminal modes. We start with the case of transverse modes directed along the axis of symmetry, for which the earlier dispersion relation is once more
\begin{equation}
\Phi(u_z,k_z)=I_0(u_z)-I_2(u_z)+2k_z=0.
\end{equation}
For weak subluminal modes, we write $u_z=\uR+i\uI$ and $k_z=\overline{k}+\delta k$, where $\overline{k}$ is the lowest-order approximation for the wavenumber, and $\delta k$ is the next correction in the weak-instability expansion, as introduced in Sec.~\ref{sec:weak_subluminal}. Using the expressions derived there, we find immediately
\begin{equation}
    \overline{k}(\uR)=-\frac{I_0(\uR)-I_2(\uR)}{2}.
\end{equation}
The corresponding growth rate is
\begin{equation}
    \uI(\uR)={\pi s_z g_{\uR}(1-\uR^2)}\Bigg/{\frac{\partial}{\partial \uR}\fint dv_z \frac{g_{v_z} (1-v_z^2)}{\uR-v_z}}.
\end{equation}
Similarly, we obtain from Eq.~\eqref{eq:approx_delta_kappa}
\begin{equation}
    \delta k(\uR)=\frac{\pi^2}{4}\frac{\partial \left[(1-\uR^2)^2 g_{\uR}^2\right]}{\partial\uR}\Bigg/
    {\frac{\partial}{\partial\uR} \fint dv_z \frac{g_{v_z} (1-v_z^2)}{\uR-v_z}}.
\end{equation}

For the longitudinal modes, the dispersion relation of Eq.~\eqref{eq:dispersion_relation_axial_kappa} is once more explicitly given as
\begin{equation}\label{eq:Phi_longitudinal}
    \Phi(u_z,k_z)=k_z^2-k_z \left[I_0(u_z)-I_2(u_z)\right]+G_1 I_0(u_z)-G_0 I_1(u_z)=0.
\end{equation}
For weak instabilities, we again write the wavevector $k_z$ associated with the real phase velocity $\uR$ as $k_z=\overline{k}+\delta k$. To lowest order, $\overline{k}$ derives from the solution of Eq.~\eqref{eq:Phi_longitudinal} where all the integrals are evaluated in principal value. For notational clarity, we will denote the integrals $I_n$ evaluated in their principal value as
\begin{equation}
    P_n=\fint dv \frac{g_v v^n}{u-v}.
\end{equation}
So, there are two branches of weakly unstable modes for which
\begin{equation}
    \overline{k}_\pm=\frac{1}{2}\left[P_0-P_2\pm \sqrt{(P_0-P_2)^2+4(G_0 P_1-G_1 P_0)}\right].
\end{equation}
For each of these branches, the approximate growth rates can now be obtained by replacing the values of $\overline{k}_\pm$ into the function $\Phi$ in Eq.~\eqref{eq:Phi_longitudinal} and using the general expressions in Sec.~\ref{sec:weak_subluminal}. Below, we will show the accuracy of these expressions in an explicit example.

\section{A worked-out example}\label{sec:benchmark_example}

\begin{figure}[b!]
    \centering
    \includegraphics[width=0.6\linewidth]{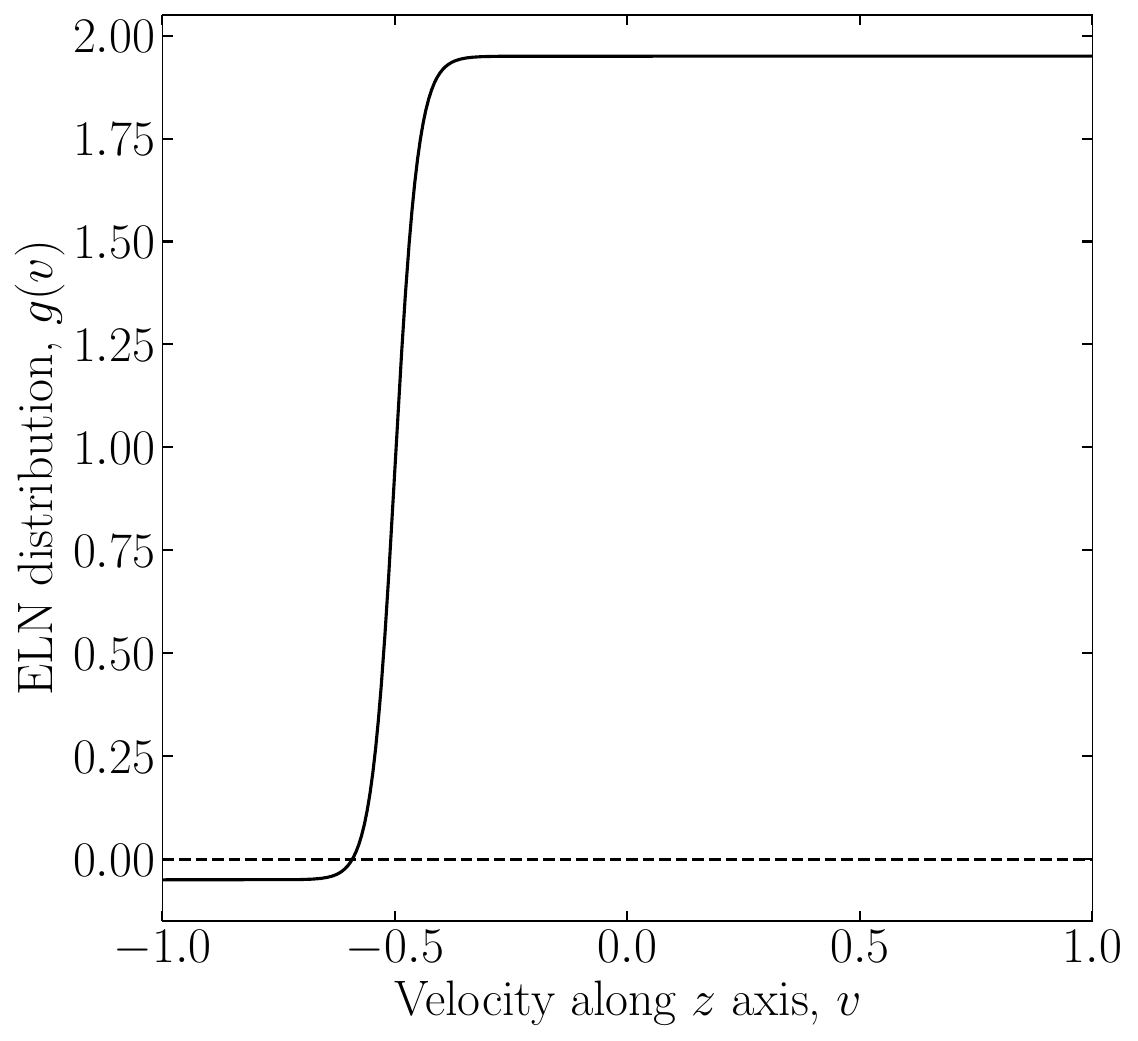}
    \caption{Angular distribution of Eq.~\eqref{eq:angular_distribution}
    chosen as a benchmark example.}
    \label{fig:angular_distribution}
\end{figure}

To illustrate these general features, we will now consider an explicit axially symmetric angular distribution and work out its regions of instability. Specifically, we focus on an angular distribution of the form
(see Fig.~\ref{fig:angular_distribution})
\begin{equation}\label{eq:angular_distribution}
    g_v=\tanh\left[20(v+0.5)\right]+0.95,
\end{equation}
which has a crossing at $v_{\rm cr}\simeq -0.59$. The shape is chosen for the crossing to be very shallow and therefore can be considered weakly unstable.

We now determine the regions of instability, i.e., the values of $\bk$ leading to unstable modes. We parameterize $\bk=\kappa\,\hat{\bn}=\kappa\,(\cos\alpha\,\hat{\bx}+\sin\alpha\,\hat{\bz})$ as before. 
There are some special values of $\alpha$ for which we can already infer some special values of $\kappa$:
\begin{itemize}
    \item for $\alpha=0$ and $\pi$, corresponding to modes directed along the axis of symmetry, the threshold values for subluminal modes to become unstable are determined by Eq.~\eqref{eq:threshold_transverse} for the two transverse modes T and T1, and by Eq.~\eqref{eq:threshold_longitudinal} for the longitudinal modes, which we call L1 and L2;
    \item for any $\alpha$, the transverse mode T is stable on the luminal sphere $u=1$, which therefore is another boundary for the region of instability;
    \item for $\cos\alpha=v_{\rm cr}$, i.e., $\alpha\simeq 126.3^\circ$, the modes on the luminal sphere have zero imaginary part, so all four solutions of the determinant equation for $\kappa$, Eq.~\eqref{eq:dispersion_kappa}, are real. This was proven in Paper~I~\cite{Fiorillo:2024bzm} (strictly speaking, we only proved that either two or four eigenvalues are real; in this case, all four of them are). Therefore, for all modes at this value of $\alpha$ there will be a boundary at $u=1$.
\end{itemize}

For all other values of $\alpha$, to identify the boundaries of the region of instability, we must proceed as discussed in Sec.~\ref{sec:boundaries_general}, namely:
\begin{itemize}
    \item for subluminal modes ($|u|<1$), we find the real values of $u$ such that one of the solutions of Eq.~\eqref{eq:dispersion_kappa} for $\kappa$ is real and positive; this will correspond to a real wavevector for which there is a subluminal mode with vanishing imaginary part, separating Landau-damped modes from growing modes;
    \item for superluminal modes ($|u|>1$), we find the real values of $u$ such that both $\Phi(u,\kappa,\bn)$ and $\partial \Phi(u,\kappa,\bn)/\partial u$ vanish simultaneously.
\end{itemize}

This strategy reveals for each $\alpha$ the values of $u_i(\alpha)$ marking the edge of an unstable region, and, equivalently, of the wavevector $\kappa_i(\alpha)$. Here the index $i$ runs through every solution of the above conditions; overall we can identify for each $\alpha$ four solutions, which continuously connect to the four solutions L1, L2, T, and T1 for $\alpha=0$. 

\begin{figure}[b!]
    \centering
    \includegraphics[width=\linewidth]{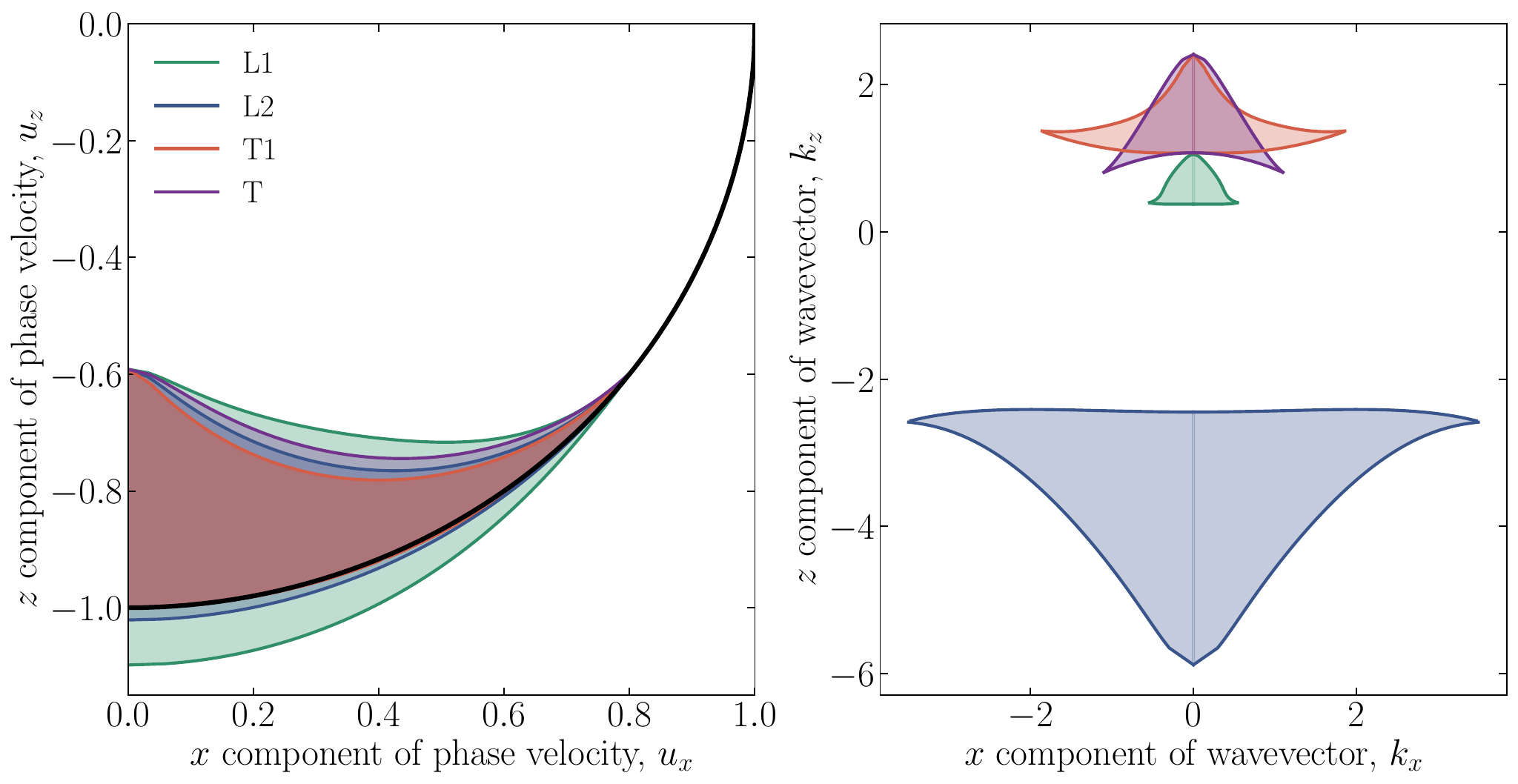}
    \caption{Regions of instability in the space of phase velocity (defined as $\bu=u\bn$, where $\bn$ is the direction of the mode and $u=\mathrm{Re}(\omega)/k$) in the left panel, and of wavevectors in the right panel. The colored regions contain an unstable mode. We mark by different colors the different families of modes identified in the main text. The luminal sphere $u=1$ is shown in solid black.}
    \label{fig:unstable_regions}
\end{figure}

\begin{figure}[b!]
    \centering
    \includegraphics[width=0.6\linewidth]{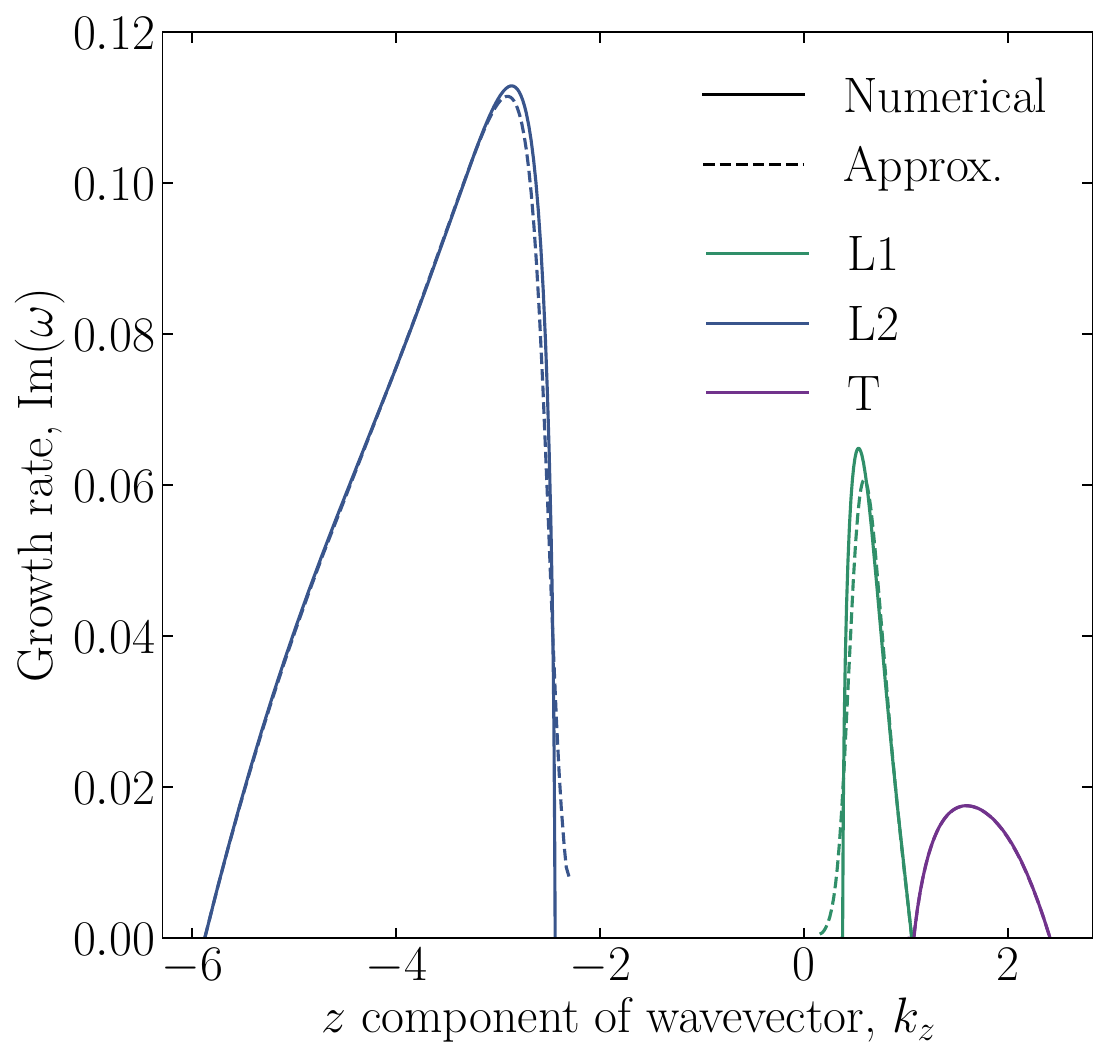}
    \vskip-6pt
    \caption{Growth rate of unstable modes as a function of wavevector. We show the exact (solid) and approximate (dashed) values of the growth rate, for both longitudinal (L1 and L2) modes, and for the two degenerate transverse modes (T).}
    \label{fig:approx_growth}
    \vskip-4pt
\end{figure}

Figure~\ref{fig:unstable_regions} shows the regions of instability both in the space of phase velocity $\bu=u(\cos\alpha,0,\sin\alpha)$ and in the space of wavevector $\bk$. These regions of instability are easiest to interpret in terms of the phase velocity. For each mode, the unstable regions open along the vertical axis ($\alpha=\pi$) at $u_z=v_{\rm cr}$ as expected, and close along the same axis at a phase velocity just above the speed of light, as we predicted qualitatively in Sec.~\ref{sec:weak_superluminal}. Along this axis, as expected, the two modes T1 and T are identical, and therefore their regions of instability touch at $\alpha=\pi$. Notice that, while all the modes are unstable in a similar range of phase velocity, their ranges of unstable wavenumbers are completely different, as they all obey different dispersion relations. Moving away from the vertical axis, the regions of instability in phase velocity space narrow down until they close at $\cos\alpha=v_{\rm cr}$ precisely on the luminal sphere, as we had anticipated. This is characteristic of distributions which are weakly unstable; as we discussed in Sec.~\ref{sec:weak_superluminal}, growth rates of order $\uI\sim 1$ can exist also when the phase velocity tends to infinity, so a generically unstable distribution might have an open region of instability in the space of phase velocities. For our benchmark example, which is close to stability, this is not the case.

We now turn to the values of the growth rates, and how they compare with the weak instability approximation introduced in Secs.~\ref{sec:weak_subluminal} and~\ref{sec:growth_axi_weak_subluminal}. We focus on modes directed along the axis of symmetry only, so that a single component of the wavevector $k_z$ is considered. In Fig.~\ref{fig:approx_growth}, we show the numerical values of the growth rates for the three family of modes, obtained by solving the exact dispersion relations Eqs.~\eqref{eq:transverse_dispersion_relation} for mode~T and~\eqref{eq:dispersion_relation_axial_kappa} for modes~L1 and~L2. The growth rates for all modes have a characteristic humped structure. At $k_z=k_\pm$, identified in Eq.~\eqref{eq:threshold_longitudinal}, the growth rate for modes L1 and L2 pass through~0, transitioning from Landau-damped to growing as their phase velocity $u_z=v_{\rm cr}$. The analogous transition for mode T is at $k_z$ identified by Eq.~\eqref{eq:threshold_transverse}. For all modes, the unstable hump closes with a characteristic square-root behavior with a superluminal phase velocity very close to the speed of light. We repeat that the clear separation of two humps for L1 and L2 is mainly due to the distribution being close to stability; for a generic distribution, there may be no superluminal boundary to the region of instability, with modes existing all the way through the point $k_z=0$ where the phase velocity is infinite, so that the two humps of L1 and L2 may merge into a single hump. This is not the case for the weakly unstable distributions we are focusing on here.

We also show in Fig.~\ref{fig:approx_growth} the approximate growth rates obtained through the weak instability procedure. They are proportional to the number of particles moving in resonance with the wave. Throughout the region of instability, the approximation is very good, confirming the validity of the resonance picture. For the transverse mode T, the numerical and approximate curves are essentially superposed. Close to the superluminal edge of each of the unstable regions, the weak instability approximation fails to reproduce the precise square-root behavior, since it is intrinsically based on a resonant picture which requires subluminal phase velocities -- for superluminal phase velocities, in the limit of infinitely sharp resonance, there are no neutrinos to resonate with. Overall, the weak instability approximation provides an excellent reproduction of the numerical growth rates, and confirms the intuitive picture we have constructed.

\section{Discussion}\label{sec:discussion}

In this series, beginning with Paper~I \cite{Fiorillo:2024bzm}, we study weakly unstable modes of the neutrino fast flavor type. One motivation is that self-consistent astrophysical settings are unlikely to produce strong instabilities, which on the contrary can be taken as signatures of inconsistent neutrino transport in traditional numerical simulations. A second motivation is that the only unstable modes that are guaranteed by a crossed angle distribution are weak ones with a wavevector pointing in the direction of a crossing line of the neutrino angle distribution. In this sense, unstable modes at the edge of instability play a special role and provide a theoretical laboratory for a deeper understanding of the phenomenon of fast flavor waves.

In Paper~I, we have laid the conceptual foundations for these weak fast-flavor instabilities, showing that they arise from resonant emission of collective flavor waves from individual neutrino modes. This wave-particle interaction is exactly analogous to the one producing several kinds of plasma instabilities. In this second paper, we have used this qualitative picture to predict the properties of weak instabilities driven by resonance. 

Following this path, we have here turned to a number of formal properties of fast flavor instabilities that had gone unnoticed. In particular, we have shown that the dispersion relation bears a strong resemblance to the one of electromagnetic waves. The collective flavor field $\psi^\mu$, the off-diagonal element of the flavor density matrix and its flux, is analogous to the electromagnetic field $A^\mu$, and obeys a transversality constraint coming from lepton number conservation $\partial_\mu \psi^\mu=0$. For the moment, we have used this formal analogy mainly as a means of simplifying certain expressions of the dispersion relation. Its deeper physical meaning is mainly related to the relativistic invariance of the theory.

Besides this formal analogy, our main results are (i)~to provide criteria to identify the regions of instability in wavevector space, and (ii)~to provide approximate expressions for the growth rate of resonant unstable modes. In doing so, we have called attention to some forgotten or underappreciated issues. In particular, we have shown that for crossed axially symmetric distributions, the guaranteed unstable modes need not be directed along the axis of symmetry. In other words, Morinaga's theorem, for which we gave an intuitive proof in Paper~I, does not apply to one-dimensional systems. Unstable modes are only guaranteed to exist when they point close to the direction of a crossing line, since they are then resonant with ``flipped'' neutrinos. On the other hand, if the axisymmetric angular distribution has a single crossing, we have shown here by a simple geometrical argument that it \textit{does} guarantee unstable modes along the axis of symmetry.

Our new criteria to identify the unstable wavenumbers generalize the findings of Capozzi et al.~\cite{Capozzi:2019lso}. In our language, this work identified the threshold values of $k_z$ for the appearance of unstable subluminal modes along the axis of symmetry of an axisymmetric angular distribution. Here we provide criteria valid for modes along an arbitrary direction, and especially show that additional threshold values of $k_z$ might arise for superluminal modes. Indeed, for weakly unstable configurations, superluminal modes are not expected to be unstable, so we identify the existence of an additional boundary to the region of instability. Most importantly, our approximate expressions for the growth rate have the same nature as the historical estimates of the growth rate of, e.g., beam-plasma instabilities. These results transcend purely mathematical interest because their structure confirms the resonance picture developed in Paper~I, i.e., the growth rate is proportional to the lepton number moving with the same phase velocity as the unstable mode. In turn, the resonant nature of weak instabilities suggests that their evolution cannot be captured by a description in terms of a few moments of the angular distributions, since the instability saturation depends on local properties of these distributions. It would therefore be interesting to understand the connection of this conclusion with the moment-based approaches to the description of the fast flavor evolution (see, e.g., Refs.~\cite{Froustey:2023skf, Froustey:2024sgz, Kneller:2024buy}), which presumably are only applicable when the instability is so strong as to become non-resonant.

In attempting a treatment of the nonlinear saturation of the instability, these approximate expressions would be key to describe the relaxation process and its typical timescales. Overall, our results here provide a link between the intuitive framework of Paper~I and the practical requirements for a description of the instability. We thus complete the construction of a physical framework for a linear theory of fast instabilities, which might serve as the basis to proceed beyond the linear regime and discuss their nonlinear evolution, the key question of interest for astrophysical applications.

The concept of fast flavor evolution assumes that neutrino masses in the form of the vacuum oscillation frequency $\omega_{\rm vac}=\Delta m^2/2E$ can be ignored relative to the dynamics entailed by angular instabilities. Several recent papers have questioned the premise that $\omega_{\rm vac}$ can be generically ignored as soon as angular instabilities develop \cite{Shalgar:2020xns, DedinNeto:2023ykt}. Indeed, weak instabilities of the type studied here may be particularly vulnerable to modifications caused by $\omega_{\rm vac}$ in the sense that different characteristic scales for these different phenomena may blur. These important questions must be left to future study.

\acknowledgments

DFGF is supported by the Alexander von Humboldt Foundation (Germany)
and, when this work was begun, was supported by the Villum Fonden (Denmark) under Project No.\ 29388 and the European Union's Horizon 2020 Research and Innovation Program under the Marie Sk{\l}odowska-Curie Grant Agreement No.\ 847523 ``INTERACTIONS.'' GGR acknowledges partial support by the German Research Foundation (DFG) through the Collaborative Research Centre ``Neutrinos and Dark Matter in Astro- and Particle Physics (NDM),'' Grant SFB-1258-283604770, and under Germany’s Excellence Strategy through the Cluster of Excellence ORIGINS EXC-2094-390783311. We thank the Galileo Galilei Institute for Theoretical Physics (Florence, Italy) for hospitality and the INFN for partial support during work on this project.

\bibliographystyle{JHEP}
\bibliography{References.bib}

\end{document}